\documentclass{article}

\usepackage{amsmath}
\usepackage{amssymb}
\usepackage{amsfonts}
\usepackage{bm}
\usepackage[dvips,xdvi]{graphicx}

\begin{document}


\title{\bf Comparison of Linear Viscoelastic Behaviors of Langevin- and Jump-Rouse Models}
\author{Takashi Uneyama \\
Department of Materials Physics, Graduate School
 of Engineering, 
Nagoya University, \\
Furo-cho, Chikusa, Nagoya 464-8603, Japan}

\maketitle

\begin{abstract}
The Rouse model with harmonic springs and the Langevin equation (Langevin-Rouse model)
is widely used to describe the linear viscoelasticity of unentangled polymer melts.
A similar model, in which the Langevin equation is replaced by discrete local jump dynamics
(the jump-Rouse model),
is also used to describe the dynamics of some systems such as
entangled polymer melts.
Intuitively, we expect that the Langevin- and jump-Rouse models give similar linear
viscoelastic behaviors. However, the Langevin- and jump-Rouse models are
not equivalent, and their linear viscoelastic behaviors can be different.
In this work, we compare the shear relaxation moduli of the Langevin- and
jump-Rouse model in detail. We develop a jump rate model in which the
resampling ratio is tunable. By using this jump rate model, we can
smoothly connect the jump-Rouse model and the Langevin-Rouse model.
We perform kinetic Monte Carlo simulations to calculate the
shear relaxation modulus data of the jump-Rouse model.
We compare the simulation results with
various numbers of beads and resampling ratios, and show that the
shear relaxation moduli of the Langevin- and jump-Rouse models are
similar but slightly different.
We analyze the short-time relaxation behavior of the
jump-Rouse model, and show that the local jumps give a single Maxwell type
relaxation in the short-time region.
\end{abstract}

%
%
%

%

\section{INTRODUCTION}

Dynamics of unentangled polymer melts in or near equilibrium can be
well described by the Rouse model\cite{Rouse-1953,Doi-Edwards-book}.
In the Rouse model, a single tagged polymer chain in the system
is considered, and the tagged chain is expressed as an array of
beads linearly connected by harmonic springs.
Then the dynamics of the tagged chain is modeled
by a simple Langevin equation (the Langevin-Rouse model).
Although the Rouse model is originally
developed as a model for dilute polymer solutions, it rather
describes the dynamics of unentangled polymer melts.
This is because the interaction between segments is screened\cite{Doi-Edwards-book,Uneyama-2021}
and the hydrodynamic interaction is negligible in polymer melts.

The application of the Rouse model is not limited to simple unentangled
polymer melts.
For example, it can be used to describe the relaxation
mode of an entangled polymers by the constraint-release (CR) mechanism\cite{Graessley-1982,Watanabe-1999}.
In the simple reptation model\cite{Doi-Edwards-book}, a single tagged
entangled polymer chain is modeled as a chain confined in a tube-like constraint,
and the tagged chain can relax by the reptation motion along the tube
and by the contour length fluctuation.
Because the tube is formed by the surrounding chains, the relaxation
of the surrounding chains induce some local motions of the tube.
Such local motions may be interpreted as the local rearrangement of tube
segments described by the Rouse-like dynamics (the CR-Rouse model)\cite{Graessley-1982}.
In the CR-Rouse model, the tube segment moves by a local jump\cite{Orwoll-Stockmayer-1969}, not by
the Langevin equation.

The dynamics of unentangled polymers with associative groups is sometimes described
by a similar Rouse-like model\cite{Baxandall-1989,Chen-Tudryn-Colby-2013,Jiang-Zhang-Tang-Yang-2020}.
Associative groups effectively work as transient cross-links.
An associative group in a cross-link
can dissociate from the cross-link by the thermal fluctuation,
and then reassociate to another cross-link.
The motion of an associative group may look as discrete jumps.
Then the chain dynamics can be modeled by the Rouse-like dynamics (the sticky Rouse model).
The dynamics of unentangled supercooled polymers\cite{Bennemann-Baschnagel-Paul-Binder-1999,Yamamoto-Onuki-2002}
may also be (roughly) described by a similar model.
In a supercooled liquid, segmental motion
becomes highly correlated and constrained (the dynamic heterogeneity)\cite{Yamamoto-Onuki-1998}.
If we observe the motion of a single particle in a supercooled liquid,
at the short-time scale it is effective trapped in a cage formed by surrounding particles.
At the long-time scale, the particle escapes from the cage and then trapped to another cage again\cite{Yamamoto-Onuki-1998,Hachiya-Uneyama-Kaneko-Akimoto-2019,Uneyama-2020}.
At the long-time scale, the segmental motion may be modeled by discrete jumps
rather than continuum Brownian motion. If we apply this picture to a
supercooled unentangled polymer chain, we will have the Rouse-like dynamics
driven by local jumps.

Naively, we expect that the chain dynamics of such a jump-based model
(the jump-Rouse model)
is similar to that of the Langevin-Rouse model.
Intuitively, the linear viscoelasticity of the jump-Rouse model 
will be approximated well by that of the Langevin-Rouse model.
However, because these two models are not equivalent, there is no
guarantee that the dynamic properties of two models are
the same even for the same parameter set.
Direct comparison of linear viscoelasticity data by
two models with the same parameter set will be required to judge
whether the linear viscoelasticity of the jump-Rouse model can
be approximated by the Langevin-Roue model or not.

The purpose of this work is to investigate the linear viscoelasticity
of the Rouse model described by the jump-Rouse model in detail.
To compare the Langevin- and jump-Rouse models, we develop a jump rate
model with a freely tunable resampling ratio.
Our model reduces to the Langevin-Rouse
model at a certain limit, and this property allows us to connect two models
smoothly and compare dynamic properties of two models directly.
We perform numerical simulations for the jump-Rouse model and
calculate the shear relaxation modulus. Then we compare the shear
relaxation moduli of the jump-Rouse model with those of the Langevin-Rouse model.
We show that the jump-Rouse model exhibits similar but slightly different
linear viscoelasticity, compared with that of the Langevin-Rouse model.
The jump-Rouse model exhibits a characteristic relaxation at
the short-time scale. At the long-time scale, the Langevin- and
jump-Rouse models give almost the same shear relaxation modulus.
We discuss the origin of the short-time relaxation mode in the 
jump-Rouse model.
Our result justifies the use of the shear relaxation moduli by the
Langevin-Rouse model as an approximate expression for that by the jump-Rouse model.

\section{MODEL}

\subsection{Langevin- and jump-Rouse models}

We consider a single tagged chain in the system.
The chain consists of $N$ beads connected linearly by
harmonic springs. We express the position of the $i$-th beads as $\bm{R}_{j}$
($i = 1,2,\dots,N$). The interaction energy of the chain is
\begin{equation}
 \label{interaction_energy}
 \mathcal{U}(\lbrace \bm{R}_{j} \rbrace) = \sum_{j = 1}^{N - 1} \frac{3 k_{B} T}{2 b^{2}} (\bm{R}_{j + 1} - \bm{R}_{j})^{2},
\end{equation}
where $k_{B}$ is the Boltzmann constant, $T$ is the temperature,
and $b$ is the segment size. The static properties of the chain
such as the equilibrium conformation is determined by the interaction
energy \eqref{interaction_energy}. However, the dynamic properties
such as the viscoelasticity is not solely determined by the interaction
energy. They depend on the dynamics model.

A simple yet useful dynamics model is the Rouse model\cite{Doi-Edwards-book}, where beads
obey the Langevin equation\cite{vanKampen-book}:
\begin{equation}
 \label{langevin_equation}
   \frac{d\bm{R}_{j}(t)}{dt}
    = - \frac{1}{\zeta} \frac{\partial \mathcal{U}(\lbrace \bm{R}_{j}(t) \rbrace)}{\partial \bm{R}_{j}(t)}
  + \bm{\xi}_{j}(t) .
\end{equation}
Here, $\zeta$ is the friction coefficient for a bead, and $\bm{\xi}_{j}(t)$ is the Gaussian
white noise which satisfies the fluctuation-dissipation relation.
The first and second order statistical moments for $\bm{\xi}_{j}(t)$ are
\begin{equation}
 \langle \bm{\xi}_{j}(t) \rangle = 0, \quad
 \langle \bm{\xi}_{j}(t) \bm{\xi}_{k}(t') \rangle = \frac{2 k_{B} T}{\zeta} \bm{1} \delta_{jk} \delta(t - t'),
\end{equation}
where $\langle \dots \rangle$ represents the statistical average and 
$\bm{1}$ is the unit tensor.
In this work, we call the Rouse model described by the Langevin equation \eqref{langevin_equation}
as ``the Langevin-Rouse model.''
The Langevin-Rouse model is known to describe the viscoelasticity of unentangled polymer melts well.
The shear relaxation modulus is given as\cite{Thurston-Morrison-1969}
\begin{equation}
 \label{relaxation_modulus_analytic}
 G(t) = \frac{\rho k_{B} T}{N} \sum_{p = 1}^{N - 1} \exp\left[ - \frac{6 k_{B} T}{\zeta b^{2}} 4 \sin^{2} \left(\frac{p \pi}{2 N}\right) t \right].
\end{equation}
Here, $\rho$ is the bead number density.

\begin{figure}[bth]
\begin{center}
 \includegraphics[height=40mm]{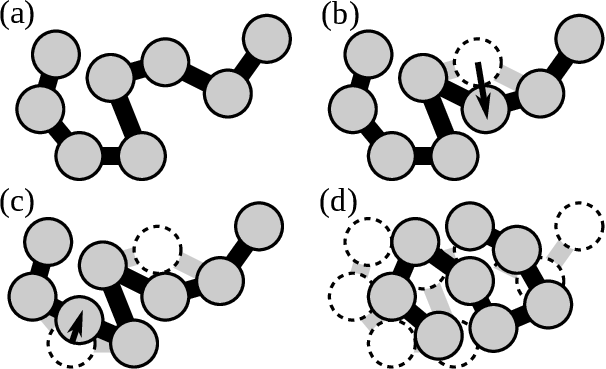}%
\end{center}
\caption{The jump-Rouse model.
 (a) The initial chain conformation. Light gray circles represent beads and
 thick black lines represent springs.
 (b) One bead in the chain jumps at the first jump event, while other beads are unchanged.
 The black arrow represents the displacement of the jumped bead.
 (c) Another bead jumps at the second jump event.
 (d) After many jump events, the whole chain conformation changes and
 the memory of the initial conformation is lost.
 \label{jump_rouse_model_image}}
\end{figure}

In the Langevin equation \eqref{langevin_equation}, beads are assumed to
move continuously in time. 
But this assumption is not always reasonable.
As we mentioned, we can interpret that the beads move rather by discrete local jumps in some systems.
We assume that discrete jumps of beads occur stochastically, and
model the dynamics by a Markovian stochastic process.
The chain conformation will relax by accumulation of jumps.
To distinguish this jump-based dynamics model from the Langevin-Rouse model,
we call this model as ``the jump-Rouse model'' in this work.
Figure~\ref{jump_rouse_model_image} shows the schematic image of the jump-Rouse model.

We describe the state of the system, or the conformation of the chain,
by the set of bead positions $\lbrace \bm{R}_{j} \rbrace$.
We introduce the transition rate from a state $\lbrace \bm{R}_{j} \rbrace$
to another state $\lbrace \bm{R}_{j}' \rbrace$ as $\Omega(\lbrace \bm{R}_{j}' \rbrace|\lbrace \bm{R}_{j} \rbrace)$.
Then the time-evolution of the time-dependent probability distribution for bead positions, $\Psi(\lbrace \bm{R}_{j} \rbrace,t)$, obeys
the following master equation\cite{vanKampen-book}:
\begin{equation}
 \label{master_equation}
  \begin{split}
   \frac{\partial \Psi(\lbrace \bm{R}_{j} \rbrace,t)}{\partial t}
   & = \int d\lbrace \bm{R}_{j}' \rbrace \,
  \big[ \Omega(\lbrace \bm{R}_{j} \rbrace|\lbrace \bm{R}_{j}' \rbrace) \Psi(\lbrace \bm{R}_{j}' \rbrace,t) 
   - \Omega(\lbrace \bm{R}_{j}' \rbrace|\lbrace \bm{R}_{j} \rbrace) \Psi(\lbrace \bm{R}_{j} \rbrace,t) \big] .
  \end{split}
\end{equation}
Here $\int d\lbrace \bm{R}_{j}' \rbrace = \int \prod_{j = 1}^{N} \bm{R}_{j}'$ is
the integral over all the possible states (or conformations).
If we assume that the jumps of beads are statistically independent,
the transition rate can be decomposed into jump rates for individual beads:
\begin{equation}
 \label{transition_and_jump_rates}
  \Omega(\lbrace \bm{R}_{j}' \rbrace|\lbrace \bm{R}_{j} \rbrace)
  = \sum_{j} \left[ \prod_{k \neq j} \delta(\bm{R}_{k}' - \bm{R}_{k}) \right]
  W_{j}(\bm{R}_{j}' | \bm{R}_{j} ,\bar{\bm{R}}_{j}).
\end{equation}
Here, $W_{j}(\bm{R}_{j}' | \bm{R}_{j} ,\bar{\bm{R}}_{j})$ is the jump rate for the $j$-th bead, and
$\bar{\bm{R}}_{j}$ represents the local equilibrium position for the
$j$-th bead:
\begin{equation}
 \bar{\bm{R}}_{j} =
 \begin{cases}
  \bm{R}_{2} & (j = 1), \\
  \displaystyle(\bm{R}_{j - 1} + \bm{R}_{j + 1}) / 2 & (2 \le j \le N - 1), \\
  \bm{R}_{N - 1} & (j = N).
 \end{cases}
\end{equation}
We naively expect that the jump-based stochastic process described by eqs~\eqref{master_equation}
and \eqref{transition_and_jump_rates} will be similar to the Langevin-Rouse dynamics
by eq~\eqref{langevin_equation}.
But eqs~\eqref{master_equation} and \eqref{transition_and_jump_rates}
are not equivalent to eq~\eqref{langevin_equation}, and
thus the Langevin- and jump-Rouse models are not equivalent, in general.

\subsection{Jump rate model}

The jump-Rouse model is not uniquely specified unless the explicit form
of the jump rate $  W_{j}(\bm{R}_{j}' | \bm{R}_{j} ,\bar{\bm{R}}_{j})$
is given. There are many different choices for the jump rate model.
The simplest model will be the stochastic resampling of the bead position
from the local equilibrium distribution.
For the $j$-th bead, the resampling is simply described as follows:
\begin{equation}
 \label{jump_rate_full_resampling}
  W_{j}(\bm{R}_{j}' | \bm{R}_{j} , \bar{\bm{R}}_{j}) = k_{\text{jump}} \Psi_{\text{leq},j}(\bm{R}_{j}' | \bar{\bm{R}}_{j}),
\end{equation}
where
$k_{\text{jump}}$ is the characteristic jump rate, and $\Psi_{\text{leq},j}(\bm{R}_{j}'| \bar{\bm{R}}_{j})$
is the local equilibrium distribution for the $j$-th bead.
The local equilibrium distribution becomes
\begin{equation}
 \label{local_equilibrium_distribution}
 \Psi_{\text{leq},j}(\bm{R}_{j}'| \bar{\bm{R}}_{j})
  = \left(\frac{3 \phi_{j}}{2 \pi b^{2}}\right)^{3/2} \exp\left[ - \frac{3 \phi_{j}}{2 b^{2}} (\bm{R}_{j}' - \bar{\bm{R}}_{j})^{2}\right] ,
\end{equation}
where $\phi_{j} = 1$ for $j = 1$ and $N$, and $\phi_{j} = 1$ for $2 \le j \le N - 1$.
Since eq~\eqref{local_equilibrium_distribution} is a Gaussian distribution, the new bead position
$\bm{R}_{j}'$ by eq~\eqref{jump_rate_full_resampling} is simply expressed as
\begin{equation}
 \label{position_after_full_resampling}
 \bm{R}_{j}' = \bar{\bm{R}}_{j} + \sqrt{\frac{b^{2}}{3 \phi_{j}}} \bm{w},
\end{equation}
with $\bm{w}$ being a Gaussian random number with $\langle \bm{w} \rangle = 0$
and $\langle \bm{w} \bm{w} \rangle = \bm{1}$.
The jump rate model \eqref{jump_rate_full_resampling}
satisfies the detailed balance condition, and thus the equilibrium
statistics is reproduced by iterating the sampling with eq~\eqref{position_after_full_resampling}.

Due to the Markovian nature, the number of jump events in a given time window
obeys the Poisson distribution\cite{vanKampen-book}, and the time between two successive jump
events obey the exponential distribution.
The total jump rate for the $j$-th bead is
\begin{equation}
 \label{total_jump_rate_full_resampling}
 \bar{W}_{j} = \int d\bm{R}_{j}' \, W_{j}(\bm{R}_{j}' | \bm{R}_{j}, \bar{\bm{R}}_{j}) = k_{\text{jump}},
\end{equation}
and the total jump rate for all the beads is $\bar{W} = \sum_{j = 1}^{N} \bar{W}_{j} = N k_{\text{jump}}$.
Therefore, the average time between successive jump events is $1 / N k_{\text{jump}}$.

In the jump model described above (eq~\eqref{jump_rate_full_resampling}, or the combination
of eqs~\eqref{position_after_full_resampling}
and \eqref{total_jump_rate_full_resampling}), the bead position is resampled from
the local equilibrium distribution, and the local conformation fully relaxes
just by one jump event. This is in contrast to the Langevin equation \eqref{langevin_equation},
by which the local conformation relaxes gradually.
We attempt to generalize the jump rate model and construct a stochastic model
in which the local conformation relaxes gradually.

Here we consider to generalize eq~\eqref{position_after_full_resampling} so that the 
position after a jump is correlated to the position before the jump.
The resampling from the local equilibrium distribution can be
interpreted as the long-time limit of the following Langevin equation for
the $j$-th beads, under the condition that all the other beads are fixed:
\begin{equation}
 \label{langevin_equation_fixed_local_equilibrium_position}
 \frac{d\bm{R}_{j}(t)}{dt} = - \frac{3 \phi_{j} k_{B} T}{\zeta b^{2}}
  [\bm{R}_{j}(t) - \bar{\bm{R}}_{j}] + \sqrt{\frac{2 k_{B} T}{\zeta}} \bm{\xi}_{j}(t).
\end{equation}
Here, $\bar{\bm{R}}_{j}$ is assumed to be constant, and $\bm{\xi}_{j}(t)$ is the Gaussian
noise which satisfies the fluctuation-dissipation relation. Eq~\eqref{langevin_equation_fixed_local_equilibrium_position}
is the Ornstein-Uhlenbeck process\cite{vanKampen-book}, and can be analytically solved:
\begin{equation}
 \label{langevin_equation_fixed_local_equilibrium_position_solution}
 \bm{R}_{j}(t)  =  \bar{\bm{R}}_{j}
 + \exp\left( - \frac{3 \phi_{j} k_{B} T}{\zeta b^{2}} t \right)
  [\bm{R}_{j}(0) - \bar{\bm{R}}_{j}]
  + \bm{\Xi}(t),
\end{equation}
with
\begin{equation}
 \bm{\Xi}(t) =
  \int_{0}^{t} dt' \, \exp\left[ - \frac{3 \phi_{j} k_{B} T}{\zeta b^{2}} (t  -t') \right]  \bm{\xi}_{j}(t). 
\end{equation}
$\bm{\Xi}(t)$ is a Gaussian noise and its first and second order moments are
\begin{align}
 & \langle \bm{\Xi}(t) \rangle = 0, \\
 & \langle \bm{\Xi}(t) \bm{\Xi}(t) \rangle
  = 
  \frac{b^{2}}{3 \phi_{j}}
\left[ 1 - \exp\left( - \frac{6 \phi_{j} k_{B} T}{\zeta b^{2}} t \right) \right] \bm{1}.
\end{align}
If we set $\bm{R}_{j}' = \lim_{t \to \infty} \bm{R}_{j}(t)$ and
$\bm{R}_{j} = \bm{R}_{j}(0)$, we have eq~\eqref{position_after_full_resampling}.
Here we consider to sample the bead position after finite time $t$, and
set $\bm{R}_{j}' = \bm{R}_{j}(t)$. We use
the beads at chain ends ($j = 1$ and $N$) as a reference,
and define
\begin{equation}
 \theta = 1 - \exp\left( - \frac{3 k_{B} T}{\zeta b^{2}} t \right)
\end{equation}
as the resampling ratio ($0 < \theta \le 1$). Then for $j = 1$ and $N$,
the position after the jump can be expressed as
\begin{equation}
 \label{position_after_partial_resampling_end}
 \bm{R}_{j}'  = \theta \bar{\bm{R}}_{j}
 + (1 - \theta) \bm{R}_{j}
  + \sqrt{\frac{b^{2}}{3} [1 - (1 - \theta)^{2}]} \bm{w},
\end{equation}
instead of eq~\eqref{position_after_full_resampling}.
For $2 \le j \le N - 1$, we have
$\exp[ - ({3 \phi_{j} k_{B} T} /{\zeta b^{2}}) t ]   =  (1 - \theta)^{2}$.
Then the position after the jump becomes
\begin{equation}
 \label{position_after_partial_resampling_center}
  \begin{split}
  \bm{R}_{j}' &  = [1 - (1 - \theta)^{2}] \bar{\bm{R}}_{j}
 + (1 - \theta)^{2} \bm{R}_{j}
   + \sqrt{\frac{b^{2}}{6} [1 - (1 - \theta)^{4}]} \bm{w}.
  \end{split}
\end{equation}
The Langevin equation \eqref{langevin_equation_fixed_local_equilibrium_position_solution}
satisfies the detailed balance condition, and thus
both eqs~\eqref{position_after_partial_resampling_end} and 
\eqref{position_after_partial_resampling_center} are detailed-balanced.
Thus they reproduce the equilibrium statistics for bead positions.
Eqs~\eqref{position_after_partial_resampling_end} and \eqref{position_after_partial_resampling_center}
correspond to the following jump rate, instead of eq~\eqref{jump_rate_full_resampling}:
\begin{equation}
 \label{jump_rate_partial_resampling}
 \begin{split}
  & W_{j}(\bm{R}_{j}'|\bm{R}_{j}|\bar{\bm{R}}_{j})  = 
  \frac{k_{\text{jump}} }{(2 \pi b^{2} \beta_{j})^{3/2}}
  \exp\Bigg[ - \frac{1}{b^{2} \beta_{j}} [\bm{R}_{j}' - \alpha_{j} \bar{\bm{R}}_{j} - (1 - \alpha_{j})^{2} \bm{R}_{j}]^{2} \Bigg],
 \end{split}
\end{equation}
with
\begin{equation}
 \label{partial_resampling_factor_alpha}
 \alpha_{j} = 
  \begin{cases}
   \theta & (j = 1, N), \\
   1 - (1 - \theta)^{2} & (2 \le j \le N - 1),
  \end{cases}
\end{equation}
\begin{equation}
 \label{partial_resampling_factor_beta}
 \beta_{j} = 
  \begin{cases}
   \displaystyle \frac{1 - (1 - \theta)^{2}}{3} & (j = 1, N), \\
   \displaystyle \frac{1 - (1 - \theta)^{4}}{6} & (2 \le j \le N - 1).
  \end{cases}
\end{equation}
The total jump rate is the same as eq~\eqref{total_jump_rate_full_resampling}: $\bar{W}_{j} = k_{\text{jump}}$
and $\bar{W} = N k_{\text{jump}}$.

According to eq~\eqref{position_after_partial_resampling_center}, the relaxation time
of the local position can be estimated as $\tau_{\text{local}} = [1 - (1 - \theta)^{2}] / k_{\text{jump}}$.
We require the thus estimated local relaxation time to be consistent with
the local relaxation time of the Langevin-Rouse model:
$\tau_{\text{local}} = \zeta b^{2} / 6 k_{B} T$.
This requirement gives the following estimate for the characteristic jump rate:
\begin{equation}
 \label{characteristic_jump_rate}
 k_{\text{jump}} = \frac{6 k_{B} T}{\zeta b^{2}} \frac{1}{1 - (1 - \theta)^{2}}.
\end{equation}
By using eq~\eqref{characteristic_jump_rate}, the time scale of the
stochastic process can be directly mapped to that of the Langevin-Rouse model.
We employ the jump rate model described by eqs~\eqref{jump_rate_partial_resampling} and
\eqref{characteristic_jump_rate} in what follows.
To be fair, we comment that our resampling model shown above is not a unique choice.
We can design other resampling models as long as they give the correct equilibrium
distribution.

At the limit of $\theta \to 0$, a displacement by a single
jump becomes infinitesimally small. Also, eq~\eqref{characteristic_jump_rate}
can be simplified as
\begin{equation}
 \label{characteristic_jump_rate_small_theta_limit}
 k_{\text{jump}} = \frac{3 k_{B} T}{\zeta b^{2}} \frac{1}{\theta}.
\end{equation}
We may interpret $k_{\text{jump}}$ in eq~\eqref{characteristic_jump_rate_small_theta_limit}
as the inverse of the effective time step size $\Delta t_{\text{eff}}$.
(The actual time step size is a stochastic quantity in this model, but
the distribution of the time step size is not important in the discussion shown below.)
Then, eqs~\eqref{position_after_partial_resampling_end} and \eqref{position_after_partial_resampling_center}
can be rewritten as follows:
\begin{equation}
 \label{position_after_partial_resampling_small_theta_limit}
  \Delta \bm{R}_{j} = 
  - \frac{3 \phi_{j} k_{B} T}{\zeta b^{2}} (\bm{R}_{j} - \bar{\bm{R}}_{j}) \Delta t_{\text{eff}} 
  + \sqrt{\frac{2 k_{B} T}{\zeta} \Delta t_{\text{eff}}} \bm{w},
\end{equation}
where $\Delta \bm{R}_{j} = \bm{R}_{j}' - \bm{R}_{j}$ is the displacement of the beads during
time step $\Delta t_{\text{eff}}$.
Eq~\eqref{position_after_partial_resampling_small_theta_limit} corresponds to
the discretized version of the Langevin equation \eqref{langevin_equation}.
Therefore, at the limit of $\theta \to 0$, our jump-Rouse model reduces to
the Langevin-Rouse model.

In the case of $\theta = 1$, 
eq~\eqref{jump_rate_partial_resampling} simply reduces to
eq~\eqref{jump_rate_full_resampling}.
Thus the simple resampling from the local equilibrium distribution is recovered.
In this case, the characteristic jump rate becomes
\begin{equation}
 k_{\text{jump}} = \frac{6 k_{B} T}{\zeta b^{2}}.
\end{equation}

%

\section{SIMULATION}

We perform simulations for the jump-Rouse model, in order to calculate
the linear viscoelasticity. In this work, we simulate the jump process
by using the kinetic Monte Carlo (kMC) method\cite{Gillespie1976,Gillespie2007}. The kMC method enables
us to simulate the stochastic process exactly.

In what follows, we use the dimensionless units. We set $k_{B} T = 1$,
$b = 1$, and $3 k_{B} T / \zeta b^{2} = 1$.
Also, we set the bead density to unity, $\rho = 1$, because the bead density
is not that important in our model. In what follows, we express all the
physical quantities in the dimensionless units.

In the kMC method, the time-evolution of the bead positions $\lbrace \bm{R}_{i}(t) \rbrace$
is described by sequence of resampling events. We describe the time where the
$i$-th resampling event occurs as $t_{i}$ ($i = 1,2,3,\dots$).
For convenience, we set $t_{0} = 0$. The resampling events occur stochastically,
and thus the time step between two successive events $t_{i} - t_{i - 1}$ is a stochastic quantity.
$t_{i} - t_{i - 1}$ obeys the exponential distribution with
$\langle t_{i} - t_{i - 1} \rangle = 1 / N k_{\text{jump}}$.
In dimensionless units, $k_{\text{jump}} = 2 / [1 - (1 - \theta)^{2}]$, and
thus $t_{i}$ is sampled as:
\begin{equation}
 t_{i} = t_{i - 1} + \frac{1 - (1 - \theta)^{2}}{2 N} \epsilon,
\end{equation}
where $\epsilon$ is a random number sampled from the exponential distribution
with the first order moment $\langle \epsilon \rangle = 1$.
At $t = t_{i}$, the target bead index is randomly selected
from $j = 1,2,\dots, N$, with the equal probability.
Then the position of the selected bead is resampled.
If we assume that the $j$-th bead is selected, its position is resampled as
\begin{equation}
 \bm{R}_{j}(t_{i + 1}) = \alpha_{j} \bar{\bm{R}}_{j}(t_{i})
  + (1 - \alpha_{j}) \bm{R}_{j}(t_{i}) + \sqrt{\beta_{j}} \bm{w},
\end{equation}
with $\alpha_{j}$ and $\beta_{j}$ defined by eqs~\eqref{partial_resampling_factor_alpha}
and \eqref{partial_resampling_factor_beta}.
$\bm{w}$ is a random number sampled from the Gaussian random number of which moments are
$\langle \bm{w} \rangle = 0$ and $\langle \bm{w}\bm{w} \rangle = \bm{1}$.
Other bead positions are unchanged by this resampling: $\bm{R}_{k}(t_{i + 1}) = \bm{R}_{k}(t_{i})$
($k \neq j$).
By iterating the resampling events, we can generate the time-dependent
bead position $\lbrace \bm{R}_{j}(t) \rbrace$.

The kMC scheme explained above is implemented on Octave (version 5.2.0)\cite{Octave-manual},
and the simulations are performed with various parameters.
We vary the number of beads per chain $N$ as $N = 5, 10, 20, 40,$ and $80$.
We also vary the resampling ratio $\theta$ as $\theta = 0.0625, 0.125, 0.25, 0.5,$ and $1$.
For all the cases, simulations with $128$ different random seeds are 
performed. Each simulation is run upto $t = 10^{5}$.
The initial bead positions $\lbrace \bm{R}_{j}(0) \rbrace$ are sampled from the equilibrium distribution.

During the simulation, the time series data of the stress tensor are stored,
and the relaxation moduli are calculated from the stress tensor data.
The single-chain stress tensor at time $t$ is calculated as
\begin{equation}
 \label{stress_tensor}
 \hat{\bm{\sigma}}(t) = \sum_{j = 1}^{N - 1} 3 [\bm{R}_{j + 1}(t) - \bm{R}_{j}(t)] [\bm{R}_{j + 1}(t) - \bm{R}_{j}(t)].
\end{equation}
The linear response theory\cite{Evans-Morriss-book,Likhtman-chapter} gives the expression of the shear relaxation modulus
in terms of the shear stress correlation:
\begin{equation}
 \label{relaxation_modulus_linear_response_theory}
 G(t) = \frac{1}{N} \langle \hat{\sigma}_{xy}(t) \hat{\sigma}_{xy}(0) \rangle_{\text{eq}}.
\end{equation}
where and $\langle \dots \rangle_{\text{eq}}$ is the equilibrium ensemble average.
To improve the statistical accuracy, we use the following formula by Likhtman\cite{Likhtman-chapter} instead of
eq~\eqref{relaxation_modulus_linear_response_theory}:
\begin{equation}
 \label{relaxation_modulus_likhtman_formula}
  \begin{split}
  G(t) & = \frac{1}{5 N}
  \Big[ \langle \hat{\sigma}_{xy}(t) \hat{\sigma}_{xy}(0) \rangle_{\text{eq}}
   + \langle \hat{\sigma}_{yz}(t) \hat{\sigma}_{yz}(0) \rangle_{\text{eq}} \\
   & \qquad + \langle \hat{\sigma}_{zx}(t) \hat{\sigma}_{zx}(0) \rangle_{\text{eq}}
  \Big] + \frac{1}{30 N}
  \Big[ \langle \hat{N}_{xy}(t) \hat{N}_{xy}(0) \rangle_{\text{eq}} \\
   & \qquad + \langle \hat{N}_{yz}(t) \hat{N}_{yz}(0) \rangle_{\text{eq}}
   + \langle \hat{N}_{zx}(t) \hat{N}_{zx}(0) \rangle_{\text{eq}}
  \Big],
  \end{split}
\end{equation}
where $\hat{N}_{\alpha\beta}(t) = \hat{\sigma}_{\alpha\alpha}(t) - \hat{\sigma}_{\beta\beta}(t)$
is the normal stress difference.
We calculate the ensemble average in eq~\eqref{relaxation_modulus_likhtman_formula}
as the averages over time and random seeds.

\begin{figure}[bth]
\begin{center}
\includegraphics[height=45mm]{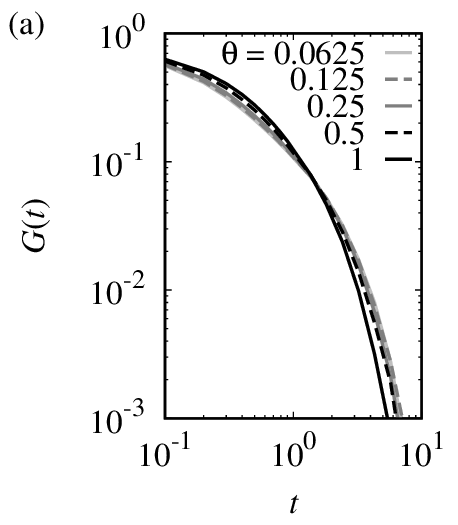}%
\includegraphics[height=45mm]{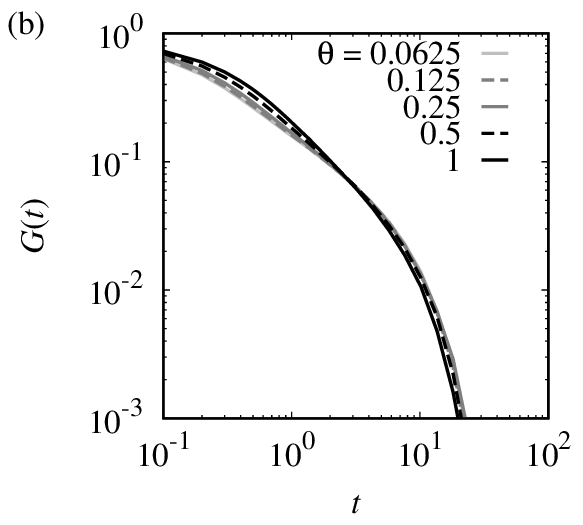}%
\includegraphics[height=45mm]{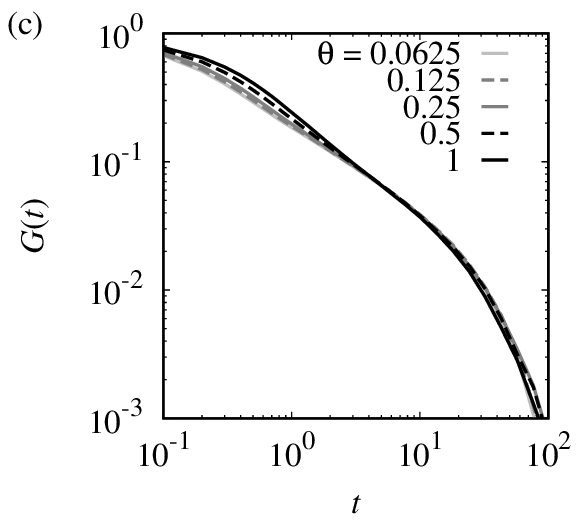}\\
\includegraphics[height=45mm]{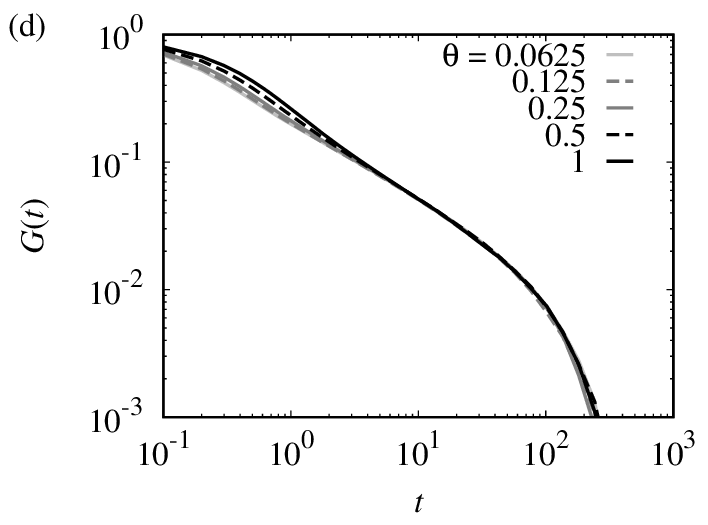}%
\includegraphics[height=45mm]{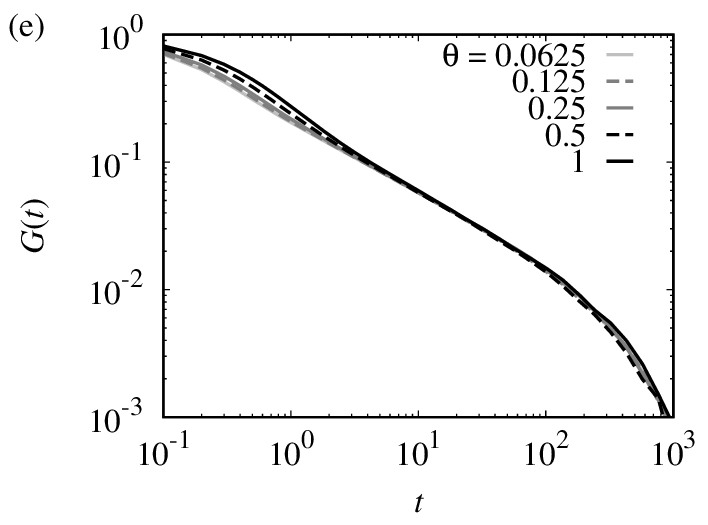}\\
\end{center}
\caption{
 Shear relaxation moduli of the jump-Rouse model for various $N$ and $\theta$.
 (a) $N = 5$, (b) $N = 10$, (c) $N = 20$, (d) $N = 40$, and (e) $N = 80$.
 \label{jump_rouse_relaxation_moduli_n}}
\end{figure} 

\section{RESULTS AND DISCUSSIONS}

\begin{figure}[bth]
\begin{center}
 \includegraphics[height=55mm]{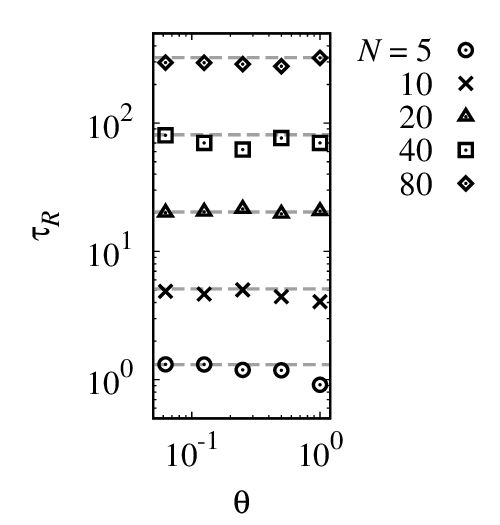}
\end{center}
\caption{
 Viscoelastic rouse times of the jump-Rouse model, $\tau_{R}$, for various $N$ and $\theta$.
 (a) $N = 5$, (b) $N = 10$, (c) $N = 20$, (d) $N = 40$, and (e) $N = 80$.
 Gray dashed lines show the viscoelastic Rouse time of the Langevin-Rouse model $\tau_{R} = 1 / 8 \sin^{2}(\pi / 2 N)$.
 \label{jump_rouse_relaxation_time}}
\end{figure}

Figure~\ref{jump_rouse_relaxation_moduli_n} shows the shear relaxation
moduli of the jump-Rouse model with various $N$ and $\theta$.
We observe that the shear relaxation modulus is not sensitive to $\theta$.
We estimate the viscoelastic Rouse time $\tau_{R}$ from the shear relaxation
modulus at the long-time region:
\begin{equation}
 \label{viscoelastic_rouse_time}
 \ln G(t) \approx - \tau_{R} / t + (\text{const}) \qquad (t \gtrsim \tau_{R}).
\end{equation}
The viscoelastic Rouse times
estimated by the least-squares fitting are shown in Figure~\ref{jump_rouse_relaxation_time}.
For comparison, the viscoelastic Rouse time for the Langevin-Rouse
model, $\tau_{R} = 1 / 8 \sin^{2}(\pi / 2 N)$ is also shown in Figure~\ref{jump_rouse_relaxation_time}.
As expected from Figure~\ref{jump_rouse_relaxation_moduli_n},
the Rouse time of the jump-Rouse model is not sensitive to $\theta$, and
almost coincides to that of the Langevin-Rouse model.
We expect that the shear relaxation modulus by the jump-Rouse model
can be well approximated by that of the Langevin-Rouse model.

\begin{figure}[bth]
\begin{center}
\includegraphics[height=50mm]{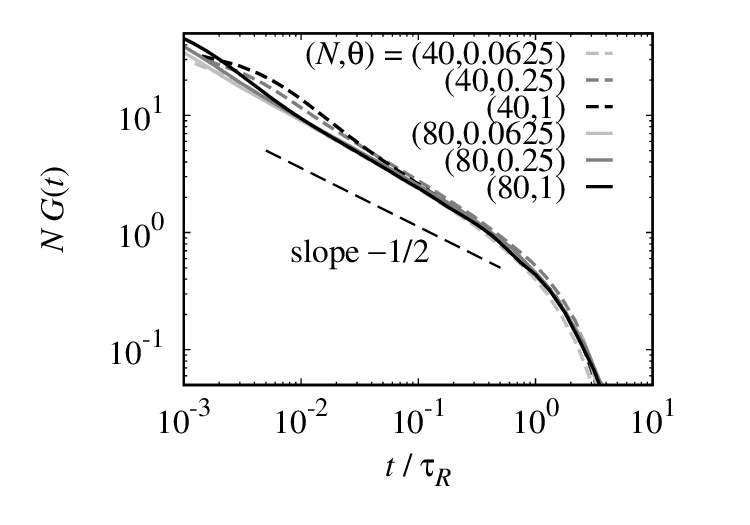}%
\end{center}
\caption{
 Reduced shear relaxation moduli of the jump-Rouse model for long chains ($N = 40$ and $80$).
 $\theta = 0.0625, 0.25,$ and $1$.
 For comparison, the power-law type relaxation with the relaxation exponent
 $1/2$ ($G(t) \propto t^{-1/2}$) is shown by a thin dashed line.
 \label{jump_rouse_reduced_relaxation_moduli}}
\end{figure}

Before we proceed to the detailed analyses for the relaxation modulus data
by the jump-Rouse model, here we discuss the terminal region. If $N$ is sufficiently
large, the Langevin-Rouse model exhibits universal relaxation behavior
which is independent of $N$.
This property makes the Langevin-Rouse model physically reasonable.
To examine whether the jump-Rouse model exhibits the universal relaxation
behavior at the terminal region or not, we reduce $t$ and $G(t)$ as
$t / \tau_{R}$ and $N G(t)$ and compare the reduced data.
Figure~\ref{jump_rouse_reduced_relaxation_moduli}
shows the reduced relaxation modulus data for $N = 40$ and $80$,
and $\theta = 0.0625, 0.25,$ and $1$.
We observe that all the data are almost collapsed. Moreover, we observe that
the power-law type relaxation behavior, $G(t) \propto N^{-1/2}$, 
the behavior well-known for the Langevin-Rouse model for sufficiently large $N$.
Thus we find that the jump-Rouse model exhibits the same universal behavior
as the jump-Rouse model, and it can be employed as a physically reasonable
model in the same way as the Langevin-Rouse model.

\begin{figure}[bth]
\begin{center}
\includegraphics[height=50mm]{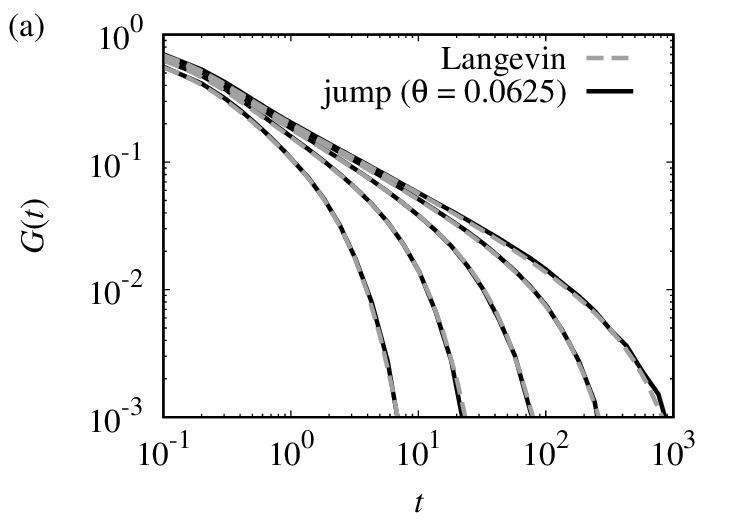}%
\includegraphics[height=50mm]{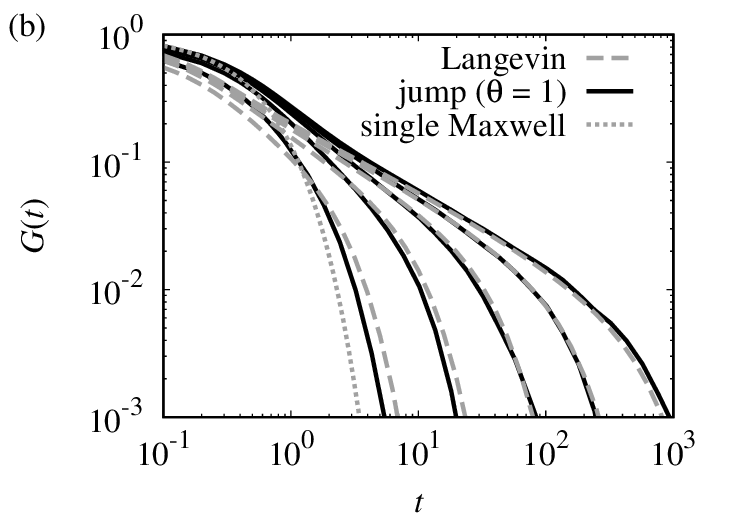}
\end{center}
\caption{
 Comparison of the shear relaxation moduli by the jump-Rouse model and the Langevin-Rouse model.
 (a) $\theta = 0.0625$ and (b) $\theta = 1$.
 $N = 5, 10, 20, 40,$ and $80$ from left to right.
 Gray dashed curves show the shear relaxation modulus by the Langevin-Rouse model (eq~\eqref{relaxation_modulus_analytic_dimensionless}).
 Gray dotted curve shows the single Maxwell model (eq~\eqref{relaxation_modulus_short_time_limit}).
 \label{jump_rouse_relaxation_moduli_theta}}
\end{figure}

If we observe the $\theta$-dependence in Figure~\ref{jump_rouse_relaxation_moduli_n} in detail,
we find that  the shear relaxation modulus changes systematically by changing $\theta$.
As increasing $\theta$, a weak shoulder appears at the short-time region ($t \lesssim 10^{0}$).
In addition, for small $N$ cases, the terminal relaxation seems to be slightly accelerated.
At the limit of $\theta \to 0$, the jump-Rouse model should reduce to
the Langevin-Rouse model, as we explained. Thus we expect that
the data for $\theta = 0.0625$ are close to the shear relaxation modulus
by the Langevin-Rouse model.
Also, we expect that the jump-Rouse model
with $\theta = 1$ is the most deviated from the Langevin-Rouse model (although
the deviation may be rather small).
Thus we compare the shear relaxation moduli
for $\theta = 0.0625$ and $1$, with those by the Langevin-Rouse model.
Figure~\ref{jump_rouse_relaxation_moduli_theta} shows
the shear relaxation moduli for $\theta = 0.0625$ and $1$, taken from
Figure~\ref{jump_rouse_relaxation_moduli_n}, and the shear relaxation modulus
of the Langevin-Rouse model. In dimensionless units, the shear relaxation modulus by eq~\eqref{relaxation_modulus_analytic}
becomes as follows:
\begin{equation}
 \label{relaxation_modulus_analytic_dimensionless}
 G(t) = \frac{1}{N} \sum_{p = 1}^{N - 1} \exp\left[ - 8 \sin^{2} \left(\frac{p \pi}{2 N}\right) t \right].
\end{equation}
We find that the data for $\theta = 0.0625$ are almost perfectly overlap
to eq~\eqref{relaxation_modulus_analytic_dimensionless}. Thus we confirm that our jump-Rouse model
reduces to the Langevin-Rouse model if $\theta$ is sufficiently small.

The data for $\theta = 1$ is deviated from the Langevin-Rouse model, and
the deviation is large at the short-time region ($t \lesssim 10^{0}$).
The jump-Rouse model exhibits a shoulder whereas the Langevin-Rouse model
exhibits a power-law type relaxation ($G(t) \propto t^{-1/2}$).
The shoulder at the short-time region can be interpreted as the relaxation
mode due to the local jumps. In the case of $\theta = 1$, the local bead position
is fully relaxed by a single jump event. Also, if the chain is sufficiently
long ($N \gg 1$), the contribution of the chain ends are negligibly small.
Therefore, we expect that we have a short-time relaxation mode which is
almost independent of $N$.
The local jumps are not correlated, and thus
there is no collective mode at the short-time scale.
This is why the jump-Rouse model does not exhibit the power-law type relaxation,
which originates from the higher order Rouse modes.


We estimate the short-time shear relaxation modulus based on
jump events.
We describe the initial bead positions as $\lbrace \bm{R}_{j} \rbrace$.
We consider the case where the $j$-th bead is resampled by the first jump.
We assume that $N$ is sufficiently large, and assume that $2 \le j \le N - 1$.
The single-chain
stress at $t = 0$ can be decomposed into the $\bm{R}_{j}$-independent and  $\bm{R}_{j}$-dependent parts:
\begin{equation}
 \label{stress_tensor_decomposed_initial}
\begin{split}
  \hat{\bm{\sigma}}(0) & = \hat{\bm{\sigma}}'_{j}(\lbrace \bm{R}_{j} \rbrace) +  3 (\bm{R}_{j + 1} - \bm{R}_{j})  (\bm{R}_{j + 1} - \bm{R}_{j})
 + 3 (\bm{R}_{j} - \bm{R}_{j - 1})  (\bm{R}_{j} - \bm{R}_{j - 1})  \\
 & = \hat{\bm{\sigma}}'_{j}(\lbrace \bm{R}_{j} \rbrace)
 + 6 (\bm{R}_{j} - \bar{\bm{R}}_{j}) (\bm{R}_{j} - \bar{\bm{R}}_{j}) 
 + \frac{3}{2} (\bm{R}_{j + 1} - \bm{R}_{j - 1})(\bm{R}_{j + 1} - \bm{R}_{j - 1}) ).
\end{split}
\end{equation}
Here, $\hat{\bm{\sigma}}'_{j}(\lbrace \bm{R}_{j} \rbrace)$ is the $\bm{R}_{j}$-independent part
of the stress tensor:
\begin{equation}
\begin{split}
 \hat{\bm{\sigma}}'_{j}(\lbrace \bm{R}_{j} \rbrace) 
 & = \sum_{k = 1}^{j - 2} 3(\bm{R}_{k + 1} - \bm{R}_{k})(\bm{R}_{k + 1} - \bm{R}_{k}) 
 + \sum_{k = j + 1}^{N - 1} 3(\bm{R}_{k + 1} - \bm{R}_{k})(\bm{R}_{k + 1} - \bm{R}_{k}).
\end{split}
\end{equation}
The stress tensor is unchanged until the first jump event, and thus $\hat{\bm{\sigma}}(t) = \hat{\bm{\sigma}}(0)$ for $t < t_{1}$.
At the first jump, only the $\bm{R}_{j}$-dependent part of the stress is changed.
If we describe the newly sampled bead position as $\bm{R}_{j}'$, the stress tensor at $t_{1} \le t < t_{2}$ becomes
\begin{equation}
 \label{stress_tensor_decomposed_first_jump}
  \begin{split}
   \hat{\bm{\sigma}}(t) & =
     \hat{\bm{\sigma}}'_{j}(\lbrace \bm{R}_{j} \rbrace)
   + 6 (\bm{R}_{j}' - \bar{\bm{R}}_{j}) (\bm{R}_{j}' - \bar{\bm{R}}_{j}) 
   + \frac{3}{2} (\bm{R}_{j + 1} - \bm{R}_{j - 1})(\bm{R}_{j + 1} - \bm{R}_{j - 1}).
  \end{split}
\end{equation}
If $t$ is sufficiently small, we will observe at most one jump event
from time $0$ to time $t$.
If there is no jump event ($t < t_{1}$), the correlation of the shear stress is simply calculated as
\begin{equation}
 \label{shear_stress_correlation_no_jump}
 \langle \hat{\sigma}_{xy}(t) \hat{\sigma}_{xy}(0) \rangle_{\text{eq}} = N - 1.
\end{equation}
If there is just one jump event ($t_{1} \le t < t_{2}$),
the correlation of the shear stress can be still analytically calculated.
In equilibrium, $j$-dependent and $j$-independent parts of the stress tensor are not correlated each other.
In addition, the $j$-dependent part can be evaluated by using the local equilibrium
distribution \eqref{local_equilibrium_distribution}:
\begin{equation}
 \label{shear_stress_correlation_single_jump}
\begin{split}
 \begin{split}
  & \langle \hat{\sigma}_{xy}(t)  \hat{\sigma}_{xy}(0) \rangle_{\text{eq}} \\
  & = (N - 3) + \Bigg\langle 
 \Bigg[ \int d\bm{R}_{j} \,  
 \Bigg[ 6 (R_{j,x} - \bar{R}_{j,x}) (R_{j,y} - \bar{R}_{j,y}) 
   \\ & \qquad
  + \frac{3}{2} (R_{j + 1,x} - R_{j - 1,x})(R_{j + 1,y} - R_{j - 1,y})\Bigg] 
  \Psi_{\text{leq},j}(\bm{R}_{j}|\bar{\bm{R}}_{j})
 \Bigg]^{2}
 \Bigg\rangle_{\text{eq}}\\
  & = (N - 3) + \frac{9}{4} \big\langle (R_{j + 1,x} - R_{j - 1,x})^{2} 
  (R_{j + 1,y} - R_{j - 1,y})^{2} \big\rangle_{\text{eq}} \\
  & = N - 2.
 \end{split}
\end{split}
\end{equation}
The probability that the $j$-th bead position is not resampled upto time $t$
is estimated as $\exp (- k_{\text{jump}} t) =  \exp(- 2 t)$. Also, the probability
to have two or more jump events will be sufficiently small and negligible.
Thus the average stress correlation becomes
\begin{equation}
 \label{shear_stress_correlation_zero_or_single_jump}
 \langle \hat{\sigma}_{xy}(t)  \hat{\sigma}_{xy}(0) \rangle_{\text{eq}}
  \approx (N - 2) + \exp( - 2 t).
\end{equation}
To derive eq~\eqref{shear_stress_correlation_zero_or_single_jump}, we
considered a jump event only for the $j$-th bead.
There are $N$ beads in the chain, and all the beads can jump with an equal
probability.
Jump of another bead reduces the stress correlation in the same way as eq~\eqref{shear_stress_correlation_zero_or_single_jump}.
By considering the jump events for all the beads,
the stress correlation is estimated as
\begin{equation}
 \langle \hat{\sigma}_{xy}(t)  \hat{\sigma}_{xy}(0) \rangle_{\text{eq}} \approx (N - 1) \exp(- 2 t).
\end{equation}
Finally we have the following simple expression for the shear relaxation modulus:
\begin{equation}
 \label{relaxation_modulus_short_time_limit}
 G(t) \approx \frac{1}{N}  \langle \hat{\sigma}_{xy}(t)  \hat{\sigma}_{xy}(0) \rangle_{\text{eq}}
  \approx \exp(- 2 t).
\end{equation}
Here we have assumed that $N$ is sufficiently large and used the approximation $(N - 1) / N \approx 1$.
Eq~\eqref{relaxation_modulus_short_time_limit} means that the shear relaxation modulus of the jump-Rouse model with $\theta = 1$
is well approximated by the single Maxwell form.
We observe that eq~\eqref{relaxation_modulus_short_time_limit} reasonably
explain the simulation data in Figure~\ref{jump_rouse_relaxation_moduli_theta}(b).
As we discussed, the jump-Rouse model exhibits universal terminal
relaxation behavior if $N$ is sufficiently large. After the short-time relaxation
mode discussed here is relaxed, we will simply observe the relaxation modulus
which is almost the same as that by the Langevin-Rouse model.

The characteristic short-time relaxation behavior by
eq~\eqref{relaxation_modulus_short_time_limit} will be, however,
not important in most of practical cases.
In the case of entangled polymers, there are various relaxation modes
such as the reptation and the contour length fluctuation modes.
The associative polymers have short-time subchain relaxation.
In the case of supercooled unentangled polymers, we have the glassy relaxation mode
at the short-time region. Then, we will observe other
relaxation modes at the time scale where the shoulder is expected to emerge.
A weak shoulder will be apparently mixed with other relaxation modes,
and then we will not observe a clear shoulder. At the long-time region,
the jump-Rouse model gives approximately the same shear relaxation modulus
as the Langevin-Rouse model. Therefore, we conclude that our simulation
result justifies the use of the shear relaxation modulus of the Langevin-Rouse
as a good approximation for that of the jump-Rouse model.

Before we end this section, we briefly discuss the temperature dependence
of the jump-Rouse model. If the temperature of the system is changed,
the relaxation time and the characteristic modulus in the Langevin-Rouse model
are changed but the relaxation mode distribution itself is not changed.
Thus the time-temperature superposition (tTS) holds for the
Langevin-Rouse model. In contrast to the Langevin-Rouse model,
the temperature dependence of the jump-Rouse model is not that clear.
The thermal activation process in the jump-Rouse model is expressed
by the resampling ratio $\theta$, and how it depends on the temperature
is not clear (while the temperature dependence of the thermal noise
$\bm{\xi}_{j}(t)$ for the Langevin-Rouse model is rather simple).
It would be natural to expect that $\theta$ decreases as the temperature
increases. Then the shape of the relaxation modulus at the short-time
region is expected to change as the temperature changes. This means
that the tTS will {\em not} hold for the jump-Rouse model.
We may be able to extract some information on the local jumps from
the failure of the tTS.
But as we discussed, the shoulder in the short-range
region will not be clearly observed in practice. If we observe only on the long-time
region, the jump-Rouse model exhibits the universal behavior which 
is independent of $\theta$. Thus we expect that the tTS will {\em approximately} hold 
for the jump-Rouse model, in most of practical cases.

\section{CONCLUSIONS}

We constructed a jump rate model with a variable resampling ratio
for the jump-Rouse model, and performed numerical simulations.
By tuning the resampling ratio, the jump-Rouse model reduces both
to the Langevin-Rouse model and the local resampling model.
We calculated the shear relaxation modulus data from by simulations,
and investigated the effects of the number of beads $N$ and the
resampling ratio $\theta$. The effect of $\theta$ on the relaxation
modulus is weak. For small $\theta$ (such as $\theta = 0.0625$), the
shear relaxation modulus by the jump-Rouse model almost coincides to
that by the Langevin-Rouse model. For $\theta = 1$, we observe small
deviations. At the short-time region where the Langevin-Rouse model exhibits
the power-law type relaxation, we observe a small shoulder in the jump-Rouse model.
Also, if $N$ is small, the viscoelastic relaxation time is slightly
accelerated in the jump-Rouse model.
We theoretically analyzed the short-time relaxation mode in the jump-Rouse model with $\theta = 1$,
and derived
that the relaxation modulus becomes a single Maxwell
model for $N \gg 1$. Except the short-time region, the shear relaxation modulus
for $\theta = 1$ is close to that by the Langevin-Rouse model.
We consider that our result justifies the use of the shear relaxation
modulus of the Langevin-Rouse model as a good approximation for
that of the jump-Rouse model.

\section*{ACKNOWLEDGMENT}
This work was supported by JST, PRESTO Grant Number JPMJPR1992, Japan,
Grant-in-Aid (KAKENHI) for Scientific Research Grant B No.~JP19H01861,
and Grant-in-Aid (KAKENHI) for Scientific Research Grant B No.~JP23H01142.


\end{document}